\documentclass[11pt,aps,nofootinbib,floatfix]{article}
\usepackage{epsfig}
\usepackage{graphicx}
\usepackage{multirow} 
\usepackage{color}
\usepackage{hyperref}
\usepackage{amsmath}
\usepackage{amssymb}
\usepackage{macros} 
\usepackage{sint_modif}   

\usepackage{subcaption}  

\newcommand\pubnumber{DAMTP-2014-87}

\def\cambridge{Department of Applied Mathematics \& Theoretical Physics,
University of Cambridge, Wilberforce Road, Cambridge CB3 0WA, United Kingdom.
}

\def\Title#1{\begin{center} {\Large #1 } \end{center}}
\def\Author#1{\begin{center}{ \sc #1} \end{center}}
\def\Address#1{\begin{center}{ \it #1} \end{center}}

\newcommand\pubblock{\rightline{\begin{tabular}{l} \pubnumber\\
         \end{tabular}}}
\newenvironment{Abstract}{\begin{quotation}  }{\end{quotation}}
\newenvironment{Presented}{\begin{quotation} \begin{center} 
             \end{center}
\bigskip 
      \begin{center}\begin{large}}{\end{large}\end{center} \end{quotation}}

\begin{document}
\begin{titlepage}
\pubblock

\vfill
\Title{
Extracting $V_{\rm us}$ from Lattice QCD simulations:\\ Recent progress and prospects}
\vfill
\Author{ Nicolas Garron} 
\Address{\cambridge}
\vfill
\begin{Abstract}
I review the current status of the determination of 
$V_{\rm us}$ from a lattice perspective.
The recent progress are very impressive: computation with $2+1$ and $2+1+1$
dynamical flavours, physical pion mass, several fine lattices, different discretisation 
of the QCD Lagrangian, etc. (see for example the plenary talk 
given by  Aida El-Khadra at this conference~\cite{Aida:2014ckm}). 
In this report, intended for non-lattice experts, I give an overview of the situation
for the computation of $f_K/f_\pi$ and $f_{+}(0)$, 
from which $V_{\rm us}$  and $V_{\rm ud}$ can be extracted. 
Besides the main features of the new computations, I also present some theoretical ideas developed in the recent years 
which allow for a cleaner determination of the relevant form factor  $f_{+}(0)$.
The experimental status has been reviewed in ~\cite{Moulson:2014cra},
and $V_{\rm us}$ from hadronic tau decay in~\cite{Passemar:CKM2014}.
\end{Abstract}
\vfill
\begin{Presented}
Proceedings of the 8th International Workshop on \\ 
the CKM Unitarity Triangle (CKM 2014), \\
Vienna, Austria, September 8-12, 2014
\end{Presented}
\vfill
\end{titlepage}
\def\thefootnote{\fnsymbol{footnote}}
\setcounter{footnote}{0}

\section{Introduction}

The Cabibbo-Kobayashi-Maskawa (CKM) matrix describes quark flavour mixing
in the Standard Model (SM).
The unitarity relation imposes for the first row
\be
\label{eq:unit}
|V_{\rm ud}|^2 + |V_{\rm us}|^2 + |V_{\rm ub}|^2  =  1  \;.
\ee
The values given by the PDG 2012 read 
\be
V_{\rm ud} = 0.97427(15)\,, \quad V_{\rm us} = 0.22534(65)\,,  \quad V_{\rm ub} = 0.00351(15) \;.
\ee
With these numerical values, 
one clearly sees why finding a deviation of Eq.(\ref{eq:unit}) 
is a difficult task, but with the constant improvement on both the experimental
and the theoretical side, the first row is a very good framework for performing 
precise tests of the SM.
We note that the  value of $ |V_{\rm ub}|^2$ is an order of magnitude smaller than the current uncertainty on 
the $|V_{\rm ud}|^2$ and $ |V_{\rm us}|^2 $, which are of the same order.\\

There is currently a huge effort in the lattice community to 
improve the determination of $V_{\rm ud}$ and $V_{\rm us}$. 
We refer the reader to FLAG~\cite{Aoki:2013ldr} for a comprehensive review.
In this proceeding, I present the recent ideas and highlight the 
newest computations.

\section{Theoretical Framework - Lattice Computation}

The basis idea is that since $ |V_{\rm us} f_+(0)| $ 
and $|V_{\rm us} /V_{\rm ud} | f_{K^\pm} / f_{\pi^\pm}$ are experimentally well measured 
(the numbers are taken from ~\cite{Aoki:2013ldr})
\bean
|V_{us} f_+(0)| &=&  0.2163(5) \\
\left| \frac {V_{us}}{V_{ud}} \right| \frac{f_{K^\pm}}{f_{\pi^\pm} } &=&  0.2758(5) \;,
\eean
one can compute $ f_+(0)$ and $ {f_{K^\pm}}/{f_{\pi^\pm}}$ on the lattice 
and extract $ V_{\rm us}$ and $V_{\rm ud}$. 
(In this report we only consider QCD in the isopsin limit $m_u=m_d$,
and therefore do not write the charge explicitly, eg $f_{p^+}=f_{p^-}$,
but electromagnetic corrections are applied~\cite{Aoki:2013ldr}.)
We start with some basic definitions of the relevant form factors, 
first the decay constant 
\be
\label{eq:defpi}
\la 0 | A_\mu | P(p) \ra = ip_\mu f_P \;, 
\qquad \mbox{ where }
A_\mu =  \bar \psi_1 \gamma_\mu \gamma_5 \psi_2 \;.
\ee 
Here $P=\bar \psi_1 \gamma_5 \psi_2$ is either a pion or a kaon, hence
Eq.~(\ref{eq:defpi}) defines $f_\pi$ and $f_K$. 
From the vector current $V_\mu =  \bar \psi_1 \gamma_\mu \psi_2$ we define
the form factors $f_+$ and $f_-$
\be
\label{eq:Vform}
\la \pi (p') | V_\mu | K(p) \ra = (p + p')_\mu f_+(q^2) + (p - p')_\mu f_-(q^2) \;,
\ee
where $q=p'-p$ is the momentum transfer. Finally we also introduce the scalar form factor $f_0$ defined by
($S = \bar \psi_1 \psi_2$ )
\be
\label{eq:Sform}
\la \pi(p') | S | K(p) \ra = \frac{m_K^2 - m_\pi^2}{m_s-m_l} f_0(q^2)  \;.
\ee
The vector Ward Identity implies a relation between the vector current and the scalar density (for non-flavour singlet)
$\partial^\mu V_\mu = (m_2 - m_1) S$. In particular, this gives
\be 
f_0(q^2) = f_+(q^2) + {q^2 \over m_K^2-m_\pi^2} f_-(q^2) \;.
\ee
In particular $f_0(0) = f_+(0)$,
hence at zero-momentum transfer the form factor can either be extracted from the vector current, Eq.~(\ref{eq:Vform}),
or from the scalar density~\cite{Na:2010uf}, see Eq.~(\ref{eq:Sform}).
A standard method introduced by~\cite{Hashimoto:1999yp} is to compute a ratio such as 
\be
\label{eq:ratio}
\frac{\la \pi | \overline s \gamma_0 l | K   \ra}{\la \pi | \overline l \gamma_0 l | \pi \ra}
\frac{\la K   | \overline l \gamma_0 s | \pi \ra}{\la K   | \overline s \gamma_0 s | K \ra}
=
{(f_0(q_{\rm max}^2))}^2 \, \frac{(m_K+m_\pi)^2}{4 m_K m_\pi}
\ee
where all the hadronic states are taken at rest and $q^2_{\rm max} = (m_K-m_\pi)^2$. 
This ratio can be numerically very well determined (most of the systematics cancel out 
and the statistical precision is better at zero-momentum).
In addition the same ratio can also be evaluated with non-vanishing momenta
(for either the pion, the kaon or both) 
and the zero-momentum transfer form factor can be obtained by an interpolation (see for example 
~\cite{Becirevic:2004ya,Boyle:2007qe}).

Simulating light quark masses is numerically expensive,
and even if nowadays physical pion masses are accessible, 
one would like to take advantage of  un-physical heavier 
quark since they are statistically
more precise. The Ademollo-Gatto theorem plays a central  
role here: the form factor $f_+(0)$ is exactly one in the $SU(3)$ flavour 
limit and the first correction is parametrised by a known function $f_2$.
In practise, one can use an Ansatz of the form:
\be
\label{eq:chipt}
f_+(0) = 1 + f_2(f,m_\pi^2,m_K^2) + \mbox{higher order}
\ee

Enormous progress have been made recently on the lattice side, 
development of new ideas, algorithms, discretisation of the Lagrangian,
and of course hardware improvement, too numerous to be explained in detail in 
this report. Instead, I highlight some important improvements developed in the last years 
relevant for the lattice computation of $V_{\rm us}$
\vspace{0.5cm}\\
{\em Theoretical developments}
\begin{itemize}
\item 
Thanks to partially twisted boundary conditions, 
the momenta are not restricted to the Fourier modes and the form factor can be computed 
directly a zero-momentum transfer~\cite{Boyle:2010bh}. 
No interpolation in momenta is required, avoiding a possible model-dependence
Ansatz.\\
\item 
The use of the scalar density (instead of the vector current) 
to extract $f_0(0)=f_+(0)$. 
One advantage is that  
in Eq~(\ref{eq:Sform}) the quantity $(m_2-m_1) S$ 
is protected by a Ward Identity and hence no renormalisation is required.
\end{itemize}
{ \em Lattice improvements}
\begin{itemize}
\item 
Simulation with physical quark masses: FNAL/MILC and RBC-UKQCD are computing 
$f_+(0)$ with light quarks down to their physical value~\cite{Bazavov:2013maa,Juettner:2014ssa}. 
FNAL/MILC simulates $2+1+1$ dynamical flavours of Highly Improved Staggered Quarks (HISQ)
and RBC-UKQCD simulates $2+1$ flavours of Domain-Wall (DW) fermions, an action notoriously 
expensive which preserves chiral-flavour symmetry at finite lattice spacing.
Hence the uncertainty due to the chiral extrapolation (which was the dominant one 
in 2013) is removed, or at least drastically reduced. 

\item Inclusion of dynamical quarks: in 2013, FLAG reported that three
collaborations (FNAL/MILC, JLQCD and RBC/UKQCD) have computed $f_+(0)$ with $2+1$ flavours, 
ie a degenerate light doublet and a strange quark in the sea. 
More recently, two collaborations (ETM and the FNAL/MILC collaborations) 
have also included a dynamical charm: although one does not expect the charm to have 
a big effect in this sector, the lattice results are becoming so precise
that this should certainly be checked. 
Let us also mention that HPQCD has computed $f_K/f_\pi$ on the $2+1+1$ MILC ensemble 
with physical quark masses~\cite{Dowdall:2013rya}.
\end{itemize}
\section{Results: 2014 update }
In 2013, the FLAG reported 
\bea
f_+(0) &=& 0.9661(32) \qquad n_f=2+1\\
f_+(0) &=& 0.9560(57)(62)  \qquad n_f=2
\eea
and noted that the major source of error came for the chiral extrapolation.
We refer the reader to the original publications for more details
\cite{Boyle:2013gsa,Bazavov:2012cd,Boyle:2010bh,Boyle:2007qe,Lubicz:2009ht,Dawson:2006qc}.

These averages do not include the most recent results which are given in 
Table~\ref{tab:details1}, together with some important features 
of the simulations.  
The action denotes the type of 
discretisation used for the Dirac operators. Even if the results 
should converge to the same continuum limit, at finite lattice 
spacing the theory suffers from distortion which are action-dependent. 
It is important to note that $f_+(0)$ is now being computed with
physical quark masses and that $2+1+1$ results are also available.
\begin{center}
\begin{table}[!h]
\begin{tabular}{ccccccl}
Collaboration & Action & $m_\pi$ (MeV) & $a$ (fm) &  $N_f$ & $f_+(0)$ & \qquad $|V_{\rm us}|$ \\
\hline
FNAL/MILC \cite{Bazavov:2013maa}  &  HISQ & $130$ & $0.06$ & $2+1+1$ & $0.9704(32)$ & $0.22290(74)(52)$ \\ 
ETM       \cite{Carrasco:2014uda} &  OS   & $210$ & $0.06$ & $2+1+1$ & $0.9683(65)$ & $0.2234(16)$\\
\end{tabular}
\caption{Summary of results for the most recent computations of $f_+(0)$, not included in the FLAG average yet.
For each computation we give the lightest simulated pion mass, the finest lattice spacing
and the number of quark flavours included in the sea. 
Note that the lightest pions mass is not necessarily simulated on the finest ensemble.
The column ``action'' corresponds to the discretisation of the Dirac operator,
see the original references for more details.
}
\label{tab:details1}
\end{table}
\end{center}
We now turn to the ratios of decay constant $f_K/f_\pi$. In their 2013 report, FLAG quoted
\bean
f_K/f_\pi &=& 1.194(5) \qquad n_f=2+1+1\\
f_K/f_\pi &=& 1.192(5) \qquad n_f=2+1\\
f_K/f_\pi &=& 1.205(6)(17) \qquad n_f=2
\eean
and again we refer the reader to the original work for more details
%
\cite{Aoki:2010dy,Aoki:2009ix,Durr:2010hr,Bazavov:2009bb,Aoki:2008sm,Allton:2008pn,Follana:2007uv,Beane:2006kx,Aubin:2004fs,Engel:2011aa,Blossier:2009bx}.
Note that some of these results were obtained with 
physical quark masses. 
At Lattice 2014, both the FNAL/MILC and the RBC-UKQCD collaborations have reported 
their new results, see Table~\ref{tab:details2}. 
\begin{table}[!h]
\begin{center}
\begin{tabular}{cccccl}
Collaboration & Action & $m_\pi$ (MeV) & $a$ (fm) &  $N_f$ & \qquad $f_K/f_\pi$ 
\\
\hline
FNAL/MILC~\cite{Bazavov:2014wgs}  &  HISQ & $130$ & $0.06$ & $2+1+1$ & $1.1956 (10)\left(^{+26}_{-18}\right)$ 
\\
RBC-UKQCD~\cite{RBC:2014tka} &  DW & $139$ & $0.08$ & $2+1$ & $1.1945(45)$  
\\
\end{tabular}
\caption{2014 Update for $f_K/f_\pi$. The details are the same as in
Table~\ref{tab:details1}. The precision is to be compared to the FLAG13 average. 
}
\label{tab:details2}
\end{center}
\end{table}

It is interesting to look at the errors in more details. For example,
for $V_{\rm us}/V_{\rm ud}$~\cite{Bazavov:2014wgs}
\be
|V_{\rm us}/V_{\rm ud}| = 0.23081(52)_{\rm LQCD}(29)_{\rm BR(K_{l2})}(21)_{\rm EM}
\qquad \mbox{ Fermilab Lattices/MILC 2014}
\ee
Even if the lattice errors still dominate, they are clearly becoming competitive

\section{Conclusions - Outlook}

The latest lattice simulations are truly impressive, 
tremendous progress have been made this year, in particular 
regarding the extraction of $f_+(0)$: the lattice simulations are now reaching the physical quark
masses and include three or four flavour of dynamical quarks. 
The errors are usually dominated by the continuum extrapolation Ansatz.
Improving this error by brute force (going to finer lattices) 
is a real challenge as it requires solving some theoretical issues (see for example~\cite{Luscher:2011kk}). 
Therefore it is very important to perform these computations with 
improved lattice actions (which have genuinely smaller lattice artifacts).
Another challenge to face is that most of the lattice simulations are done 
in ``pure'' QCD in the isospin limit $m_u=m_d$. The results are now 
so precise that the effects of this approximation are becoming visible. 
For $f_{+}(0)$ or $f_K/f_\pi$, a (model dependent) correction is applied {\em a posteriori} 
to the lattice results \cite{Aoki:2013ldr}. However, important progress have been made 
recently in that respect: see~\cite{Borsanyi:2014jba} for an implementation 
of QCD+QED at order $\alpha$ and see~\cite{Portelli:CKM2014} for a review.

\vspace{0.5cm}
{\bf Acknowledgements -} I would like to thank the organisers of CKM2014 and in particular
the conveners of WG1, Stefan Bae{\ss}ler, Anze Zupanc and Elvira G\'amiz
for their kind invitation.
I acknowledge support from STFC under the grant ST/J000434/1
and from the EU Grant Agreement 238353 (ITN STRONGnet).

\bibliography{biblio}{}
\bibliographystyle{h-elsevier}
 
\end{document}